\documentclass{ptephy_v1}
\usepackage{amsmath}
\usepackage{amssymb}
\usepackage{graphicx}
\usepackage{bm}
\usepackage{dcolumn}
\usepackage{multirow}
\usepackage{lscape}
\usepackage{mathrsfs}
\usepackage{mathtools}
\usepackage[T1]{fontenc}
\def\ve#1{{\bm{#1}}}
\def\nuc#1#2#3{{}^{#2}_{#3}\mathrm{#1}}
\def\urm#1{\scriptstyle{\text{\textrm{\textmd{\textup{#1}}}}}}

\let\temp\epsilon
\let\epsilon\varepsilon
\let\varepsilon\temp
\let\temp\relax
\let\temp\phi
\let\phi\varphi
\let\varphi\temp
\let\temp\relax
\begin{document}
\title{Relativistic correction of the Coulomb interaction in the local density approximation for energies and radii in doubly-magic nuclei}

\author[1, 2]{Tomoya Naito}
\affil[1]{
  RIKEN Interdisciplinary Theoretical and Mathematical Sciences (iTHEMS),
  Wako 351-0198, Japan}
\affil[2]{
  Department of Physics, Graduate School of Science, The University of Tokyo,
  Tokyo 113-0033, Japan
  \email{tnaito@ribf.riken.jp}}
\date{Received: \today}
\begin{abstract}
  Effects of the relativistic correction of the Coulomb interaction on doubly-magic nuclei are discussed with Skyrme Hartree--Fock calculations.
  The relativistic correction is treated by using the local density approximation.
  It is found that the correction to the total energy is about $ 2.4 \, \mathrm{MeV} $ for $ \nuc{Pb}{208}{} $, while proton and neutron radii do not change significantly.
  This difference is larger than the difference of the Coulomb exchange (Fock) energy
  calculated with the local density (Hartree--Fock--Slater) approximation
  and
  that with the exact treatment and the neutron finite-size effect.
  Effects of the correction are also compared to the correction due to the vacuum polarization.
  It is shown that the two contributions to the total energy are comparable in light nuclei, but the latter dominates in heavy nuclei,
  while the contribution of the relativistic correction to the total energy is non-negligible compared to the target accuracy of the DFT calculation.
\end{abstract}
\subjectindex{D10, D12}
\maketitle
%
%
%
%
%
%
\section{Introduction}
\par
Atomic nuclei are composed of protons and neutrons,
which interact via the nuclear and electromagnetic (EM) interactions.
Since the nuclear interaction is much stronger than the EM one,
the former determines most properties of the atomic nuclei.
Nevertheless, the EM interaction plays an important role when one focuses on nuclear properties related to the isospin symmetry breaking~\cite{
  Okamoto1964Phys.Lett.11_150,
  Friedman1971Phys.Lett.B35_543,
  Shlomo1982Phys.Scr.26_280,
  Suzuki1993Phys.Rev.C47_R1360,
  Kaneko2012Phys.Rev.Lett.109_092504,
  Baczyk2018Phys.Lett.B778_178,
  Roca-Maza2018Phys.Rev.Lett.120_202501,
  Baczyk2019J.Phys.G46_03LT01,
  Loc2019Phys.Rev.C99_014311,
  Dong2019Phys.Rev.C99_014319}.
For more details, see reviews, e.g., Ref.~\cite{
  Auerbach1983Phys.Rep.98_273}.
\par
The Hartree--Fock method or the density functional theory (DFT)~\cite{
  Hohenberg1964Phys.Rev.136_B864,
  Kohn1965Phys.Rev.140_A1133,
  Kohn1999Rev.Mod.Phys.71_1253}
is a powerful method to calculate quantum many-fermion problems,
including atomic nuclei.
In DFT, an energy density functional (EDF), which contains information on interactions, governs the accuracy of calculation.
Although the Coulomb interaction is rather simple,
first-principle EDFs, even for electronic systems,
have been derived approximately using in the local density approximation (LDA)~\cite{
  Vosko1980Can.J.Phys.58_1200,
  Perdew1981Phys.Rev.B23_5048,
  Yokota2021Phys.Rev.Research3_L012015},
and effects of the density gradient have been included empirically or phenomenologically~\cite{
  Perdew1986Phys.Rev.B33_8800,
  Perdew1992Phys.Rev.B46_6671,
  Perdew1996Phys.Rev.Lett.77_3865}.
In nuclear physics, the situation is more complicated that the nuclear interaction in medium is still under discussion,
and the derivation of \textit{ab initio} nuclear EDF is one of the current important issues~\cite{
  Drut2010Prog.Part.Nucl.Phys.64_120,
  Dobaczewski2016J.Phys.G43_04LT01,
  Yokota2019Phys.Rev.C99_024302,
  Salvioni2020J.Phys.G47_085107,
  Accorto2021Phys.Rev.C103_044304,
  Marino2021Phys.Rev.C104_024315,
  Naito2022Phys.Rev.C105_L021304}.
Therefore, most nuclear EDFs have been constructed phenomenologically with fitting to experimental data
and different EDFs may give different results~\cite{
  Li2008Phys.Rep.464_113,
  Danielewicz2009Nucl.Phys.A818_36,
  Danielewicz2014Nucl.Phys.A922_1}.
The accuracy of total energies obtained by widely used EDFs ranges approximately $ 1 $--$ 5 \, \mathrm{MeV} $~\cite{
  Stoitsov2003Phys.Rev.C68_054312,
  Kortelainen2010Phys.Rev.C82_024313},
while those of empirical mass formulae and machine learning ranges $ 100 $--$ 500 \, \mathrm{keV} $~\cite{
  Wang2014Phys.Lett.B734_215,
  Niu2018Phys.Lett.B778_48,
  Niu2018Sci.Bull.63_759,
  Yamaguchi2021Prog.Part.Nucl.Phys.120_103882},
which almost matches the desired accuracy for the study of $ r $-process nucleosynthesis~\cite{
  Lunney2003Rev.Mod.Phys.75_1021}.
\par
To fit the parameters of nuclear EDF properly, the EM contribution to the nuclear properties should be subtracted from experimental data properly.
Motivated by this issue,
in a series of works~\cite{
  Naito2018Phys.Rev.C97_044319,
  Naito2019Phys.Rev.C99_024309,
  Naito2020Phys.Rev.C101_064311},
accurate treatment of the Coulomb interaction in a Skyrme Hartree--Fock (SHF) calculation~\cite{
  Vautherin1972Phys.Rev.C5_626},
one of the branches of the nuclear DFT~\cite{
  Meng2006Prog.Part.Nucl.Phys.57_470,
  Niksic2011Prog.Part.Nucl.Phys.66_519,
  Liang2015Phys.Rep.570_1,
  Nakatsukasa2016Rev.Mod.Phys.88_045004,
  Colo2020Adv.Phys.X5_1740061},
was proposed.
In the series of works, 
the generalized gradient approximation (GGA) for the Coulomb exchange energy was introduced
to reproduce the exact Fock energy with low numerical costs~\cite{
  Naito2018Phys.Rev.C97_044319,
  Naito2019Phys.Rev.C99_024309}.
On top of it,
more higher-order contributions were taken into account; 
the spatial charge distributions of nucleons and the vacuum polarization between two protons~\cite{
  Auerbach1983Phys.Rep.98_273,
  Samaddar1986Nucl.Phys.A451_160,
  Chamel2009Phys.Rev.C80_065804,
  Roca-Maza2018Phys.Rev.Lett.120_202501,
  Naito2020Phys.Rev.C101_064311}.
From the point of view of the quantum field theory,
the finite-size effects, i.e., effects originating from the spatial charge distributions of nucleons,
corresponds to the correction of vertices
and
the vacuum polarization corresponds to the higher-order correction of the Coulomb interaction in terms of the coupling constant $ \alpha \approx 1/137 $ (the fine-structure constant).
Although these effects are just corrections of the Coulomb interaction,
which itself is already subdominant for the nuclear total energies,
it was shown that they are non-negligible compared to the desired accuracy;
for instance, the proton finite-size effect, the neutron one, and the vacuum polarization contribute to the total energy of $ \nuc{Pb}{208}{} $
in $ -8.2 \, \mathrm{MeV} $, $ +1.2 \, \mathrm{MeV} $, and $ +3.7 \, \mathrm{MeV} $, respectively~\cite{
  Naito2020Phys.Rev.C101_064311}.
\par
Then, it is natural to consider the corrections with the same order with respect to $ \alpha $;
that is, the finite-light-speed correction,
the self-energy, and 
the two-photon exchange process.
Although the self-energy contributes to the energy of electrons in atoms in the same magnitude as the vacuum polarization,
while it may be weak for the energy of nucleons in atomic nuclei 
since the nucleon mass is heavy;
the two-photon exchange may also be weak since it is also inverse proportional to the nucleon mass~\cite{
  Wiringa1995Phys.Rev.C51_38}.
In contrast, the finite-light-speed effect does not depend on the nucleon mass,
and it is known to be non-negligible in atomic structure~\cite{
  Saue2011ChemPhysChem12_3077,
  Naito2020J.Phys.B53_215002}.
Thus, in this work, the finite-light-speed correction of the Coulomb interaction is considered in the nuclear structure calculation
in order to investigate whether its effect is small.
Even if its effect is small, it is worthwhile to confirm it.
\par
The finite-light-speed correction of the Coulomb interaction is often referred to as the relativistic correction~\cite{
  Saue2011ChemPhysChem12_3077,
  Pershina2015Nucl.Phys.A944_578,
  Pershina2015Nucl.Phys.A944_578,
  Eliav2015Nucl.Phys.A944_518,
  Schwerdtfeger2015Nucl.Phys.A944_551}.
The Coulomb interaction is instantaneous,
while the velocity of photons, which mediate the EM interaction, is finite.
Once the finite light speed is considered,
the Coulomb interaction should be, therefore, corrected,
and this correction is known as the Breit correction~\cite{
  Breit1929Phys.Rev.34_553,
  Breit1930Phys.Rev.36_383,
  Breit1932Phys.Rev.39_616}.
In atomic physics, the Breit correction dominates more than the vacuum polarization of the Coulomb field formed by the atomic nucleus~\cite{
  Eides2001Phys.Rep.342_63},
and thus the Breit correction should be considered once the vacuum polarization is considered.
In nuclear physics, the vacuum polarization discussed in Ref.~\cite{
  Naito2020Phys.Rev.C101_064311}
originates from the interaction between two protons,
which is different from that in atomic physics.
Hence, it is interesting to compare which contribution dominates to the nuclear total energy,
the Breit correction or the vacuum polarization.
\par
Another interesting point is a comparison between the Breit correction and the correction due to the density gradient,
i.e., the difference between the LDA Coulomb energy and the GGA Coulomb energy.
The Breit correction had been taken into account for an exchange~\cite{
  MacDonald1979J.Phys.C12_2977}
or exchange-correlation~\cite{
  Kenny1996Phys.Rev.Lett.77_1099}
EDF in the LDA level,
and was already taken into account fully for atomic DFT~\cite{
  Naito2020J.Phys.B53_215002}.
It was found in Ref.~\cite{
  Naito2020J.Phys.B53_215002}
that the Breit correction contributes to the total energies of atoms in almost the same magnitude as but the opposite to the density gradient effect.
In the context of nuclear physics, 
it was shown in Refs.~\cite{
  LeBloas2011Phys.Rev.C84_014310,
  Roca-Maza2016Phys.Rev.C94_044313,
  Naito2019Phys.Rev.C99_024309}
that difference between the Coulomb exchange energy calculated by the exact-Fock term or GGA
and that in LDA is non-negligible in nuclear binding energies.
Thus, it should be tested to what extent the Breit correction contributes to nuclear binding energies.
\par
The covariant density functional theory, the nuclear DFT in the relativistic scheme,
is based on the covariant Lagrangian with nucleon, meson, and photon fields.
Hence, effects beyond the Coulomb interaction, such as magnetic interaction and retardation, have often been considered~\cite{
  Meng2006Prog.Part.Nucl.Phys.57_470,
  Niksic2011Prog.Part.Nucl.Phys.66_519,
  Niu2013Phys.Rev.C87_037301,
  Gu2013Phys.Rev.C87_041301}.
Therefore, it is interesting to take into account the relativistic correction of the Coulomb interaction in non-relativistic SHF calculations.
\par
The relativistic correction of the Coulomb interaction was already discussed in Refs.~\cite{
  Gomez1973Phys.Lett.B44_231,
  Shlomo1976Phys.Lett.B60_244}.
As discussed later, the correction used in these works is the same as the non-relativistic reduced form of the Breit correction for electronic systems~\cite{
  Itoh1965Rev.Mod.Phys.37_159,
  Kenny1995Phys.Rev.A51_1898},
while anomalous magnetic moments of nucleons are, of course, considered.
The aim of this paper is modern calculation of Refs.~\cite{
  Gomez1973Phys.Lett.B44_231,
  Shlomo1976Phys.Lett.B60_244}:
Deriving the EDF for such corrections
and
performing the self-consistent calculation.
I shall discuss contributions to proton and neutron radii, total energies, and single-particle energies.
The LDA is used to consider the relativistic correction;
some terms of the correction vanish in the approximation, as will be discussed.
Note that it seems that ``the Breit correction'' refers to the whole terms of the non-relativistic reduced form of the correction [Eq.~\eqref{eq:Breit_NR}] in the context of atomic physics,
while it refers to only a term of the correction in the context of nuclear physics.
In order to avoid any confusion,
hereinafter, I will use a term ``the relativistic correction'', instead of ``the Breit correction''.
\par
This paper is organized as follows:
In Sec.~\ref{sec:framework}, the theoretical framework for the relativistic correction will be shown.
In Sec.~\ref{sec:analytical}, the relativistic correction to the total energy will be estimated analytically,
and 
in Sec.~\ref{sec:SHF}, numerical results will be shown.
Finally, in Sec.~\ref{sec:conclusion}, conclusion and perspectives will be given.

%
%
%
%
%
\section{Theoretical Framework}
\label{sec:framework}
\par
In this section, I show the theoretical framework for the relativistic correction and the DFT.
Throughout this paper, an unit with $ 4 \pi \epsilon_0 = 1 $ is used
and $ c $ denotes the speed of light.
\par
Quantum many-body systems can be described by using the Hamiltonian
\begin{equation}
  \label{eq:Hamil}
  H
  =
  T
  +
  V_{\urm{ext}}
  +
  V_{\urm{int}},
\end{equation}
where $ T $, $ V_{\urm{ext}} $, and $ V_{\urm{int}} $ denote
the kinetic energy operator,
an external potential,
and
an interaction, respectively.
In the case of electronic systems,
$ V_{\urm{int}} $ is identical to the EM interaction $ V_{\urm{EM}} $,
while
in the case of atomic nuclei,
$ V_{\urm{int}} $ includes
a nuclear interaction $ V_{\urm{nucl}} $
and 
the EM interaction $ V_{\urm{EM}} $
with $ V_{\urm{ext}} \equiv 0 $.
In most works, the Coulomb interaction between protons
\begin{equation}
  \label{eq:Coul}
  V_{\urm{Coul}} \left( \ve{r}_j, \ve{r}_k \right)
  =
  \frac{e^2}{r_{jk}}
\end{equation}
is used for $ V_{\urm{EM}} $,
where $ \ve{r}_j $ and $ \ve{r}_k $ are the spatial coordinates of the protons $ j $ and $ k $,
$ \ve{r}_{jk} = \ve{r}_j - \ve{r}_k $,
and
$ r_{jk} = \left| \ve{r}_{jk} \right| $.
According to the quantum electrodynamics,
the leading-order instantaneous interaction,
in which a photon mediates between two particles with $ c = \infty $,
is the Coulomb interaction.
\subsection{Breit correction for electronic systems}
\label{subsec:Breit}
\par
In electronic systems, the next-leading-order interaction ($ O \left( 1/c^2 \right) $) in the Coulomb gauge
is called the Breit correction, whose form is~\cite{
  Breit1929Phys.Rev.34_553,
  Breit1930Phys.Rev.36_383,
  Breit1932Phys.Rev.39_616,
  Gorceix1988Phys.Rev.A37_1087,
  Saue2011ChemPhysChem12_3077}
\begin{equation}
  \label{eq:Breit}
  V_{\urm{Breit}} \left( \ve{r}_j, \ve{r}_k \right)
  =
  -
  \left[
    \frac{c \ve{\alpha}_j \cdot c \ve{\alpha}_k}{2c^2 r_{jk}}
    +
    \frac{\left( c \ve{\alpha}_j \cdot \ve{r}_{jk} \right) \left( c \ve{\alpha}_k \cdot \ve{r}_{jk} \right)}{2c^2 r_{jk}^3}
  \right],
\end{equation}
where $ \alpha $ is $ 4 \times 4 $ matrix defined by 
\begin{equation}
  \ve{\alpha}
  =
  \left(
    \begin{pmatrix}
      O_2 & \sigma_x \\
      \sigma_x & O_2 
    \end{pmatrix},
    \begin{pmatrix}
      O_2 & \sigma_y \\
      \sigma_y & O_2 
    \end{pmatrix},
    \begin{pmatrix}
      O_2 & \sigma_z \\
      \sigma_z & O_2 
    \end{pmatrix}
  \right)
\end{equation}
with the Pauli matrices $ \sigma_x $, $ \sigma_y $, and $ \sigma_z $
and the $ 2 \times 2 $ zero matrix $ O_2 $.
In isolated atoms, the vacuum polarization of the Coulomb field appears in higher order than the Breit correction~\cite{
  Eides2001Phys.Rep.342_63}.
The Breit correction includes the magnetic (current-current) interaction and the retardation in the Coulomb gauge~\cite{
  Gorceix1988Phys.Rev.A37_1087}.
In the Lorenz (covariant) gauge,
the next-leading-order interaction corresponds to 
the magnetic interaction and the retardation~\cite{
  Sakurai1967AdvancedQuantumMechanics_AddisonWesley,
  Gorceix1988Phys.Rev.A37_1087,
  Lindroth1989Phys.Rev.A39_3794}.
\par
After the Foldy-Wouthuysen-Tani transformation~\cite{
  Pryce1948Proc.RoyalSoc.Lond.A195_62,
  Foldy1950Phys.Rev.78_29,
  Tani1951Prog.Theor.Phys.6_267,
  Foldy1952Phys.Rev.87_688},
the Breit correction [Eq.~\eqref{eq:Breit}] in the non-relativistic scheme is obtained as~\cite{
  Itoh1965Rev.Mod.Phys.37_159,
  Kenny1995Phys.Rev.A51_1898}
\begin{subequations}
  \label{eq:Breit_NR}
  \begin{align}
    \tilde{V}_{\urm{Breit}} \left( \ve{r}_j, \ve{s}_j, \ve{r}_k, \ve{s}_k \right)
    & = 
      -
      \frac{\pi \hbar^2 e^2}{m_e^2 c^2}
      \delta \left( \ve{r}_{jk} \right)
      \label{eq:Breit_NR_Darwin} \\
    & \quad
      -
      \frac{e^2}{2 m_e^2 c^2}
      \ve{p}_j 
      \cdot
      \left(
      \frac{1}{r_{jk}}
      +
      \frac{\ve{r}_{jk} \ve{r}_{jk}}{r_{jk}^3}
      \right)
      \cdot
      \ve{p}_k
      \label{eq:Breit_NR_retar} \\
    & \quad
      -
      \frac{8 \pi \hbar^2 e^2}{3 m_e^2 c^2}
      \delta \left( \ve{r}_{jk} \right)
      \ve{s}_j \cdot \ve{s}_k
      \label{eq:Breit_NR_sm_1} \\
    & \quad
      -
      \frac{\hbar^2 e^2}{m_e^2 c^2}
      \ve{s}_j
      \cdot
      \left(
      \frac{3 \ve{r}_{jk} \ve{r}_{jk}}{r_{jk}^5}
      -
      \frac{1}{r_{jk}^3}
      \right)
      \cdot
      \ve{s}_k
      \label{eq:Breit_NR_sm_2} \\
    & \quad
      +
      \frac{\hbar^2 e^2}{m_e^2 c^2}
      \frac{1}{r_{jk}^3}
      \ve{s}_j
      \cdot
      \left[
      \ve{r}_{jk}
      \times
      \left(
      2 \ve{p}_k - \ve{p}_j
      \right)
      \right],
      \label{eq:Breit_NR_so}
  \end{align}
\end{subequations}
where $ m_e \approx 0.511 \, \mathrm{MeV} / c^2 $ is the electron mass~\cite{
  Zyla2020Prog.Theor.Exp.Phys.2020_083C01}
and $ \ve{p}_j $ is the momentum of the particle $ j $.
The first [Eq.~\eqref{eq:Breit_NR_Darwin}] and second [Eq.~\eqref{eq:Breit_NR_retar}]
terms correspond to
the Darwin term for the electron-electron Coulomb interaction,
which is related to Zitterbewegung~\cite{
  Sakurai1967AdvancedQuantumMechanics_AddisonWesley},
and the retardation of the Coulomb interaction, respectively;
the third and fourth terms
[Eqs.~\eqref{eq:Breit_NR_sm_1} and \eqref{eq:Breit_NR_sm_2}]
correspond to the spin-magnetic interactions;
and 
the fifth term
[Eq.~\eqref{eq:Breit_NR_so}]
corresponds to the electron-electron spin-orbit interaction~\cite{
  Itoh1965Rev.Mod.Phys.37_159,
  Kenny1995Phys.Rev.A51_1898,
  Kenny1996Phys.Rev.Lett.77_1099}.
Thus, in the Schr\"{o}dinger scheme,
if one is eager to consider the Breit correction,
Eq.~\eqref{eq:Breit_NR} should be considered on top of the Coulomb interaction
[Eq.~\eqref{eq:Coul}],
i.e.,
\begin{equation}
  \label{eq:EM}
  V_{\urm{EM}}
  =
  \frac{1}{2}
  \sum_{j, \, k}
  \left[
    V_{\urm{Coul}} \left( \ve{r}_j, \ve{r}_k \right)
    +
    \tilde{V}_{\urm{Breit}} \left( \ve{r}_j, \ve{s}_j, \ve{r}_k, \ve{s}_k \right)
  \right],
\end{equation}
which will be referred to as the Coulomb--Breit interaction.
\subsection{Beyond Coulomb interaction for nuclear systems}
\label{subsec:Breit_nucl}
\par
Corrections beyond the Coulomb interaction for nuclear systems were discussed several decades ago~\cite{
  Gomez1973Phys.Lett.B44_231,
  Shlomo1976Phys.Lett.B60_244}.
The correction was
\begin{subequations}
  \label{eq:Breit_nucl}
  \begin{align}
    \tilde{V}_{\urm{Rel}} \left( \ve{r}_j, \ve{s}_j, \ve{r}_k, \ve{s}_k \right)
    & =
      - 
      \frac{e_{\tau_j} e_{\tau_k}}{2 M^2 c^2}
      \ve{p}_j 
      \cdot
      \left(
      \frac{1}{r_{jk}}
      +
      \frac{\ve{r}_{jk} \ve{r}_{jk}}{r_{jk}^3}
      \right)
      \cdot
      \ve{p}_k
      \label{eq:Breit_nucl_retar} \\
    & \quad
      -
      \frac{2 \hbar \mu_{\tau_j} e_{\tau_k}}{M c^2}
      \frac{1}{r_{jk}^3}
      \ve{s}_j
      \cdot
      \left( \ve{r}_{jk} \times \ve{p}_k \right)
      +
      \frac{2 \hbar \mu_{\tau_k} e_{\tau_j}}{M c^2}
      \frac{1}{r_{jk}^3}
      \ve{s}_j
      \cdot
      \left( \ve{r}_{jk} \times \ve{p}_j \right)
      \label{eq:Breit_nucl_so} \\
    & \quad
      -
      \frac{4 \mu_{\tau_j} \mu_{\tau_k}}{c^2}
      \ve{s}_j
      \cdot
      \left(
      \frac{3 \ve{r}_{jk} \ve{r}_{jk}}{r_{jk}^5}
      -
      \frac{1}{r_{jk}^3}
      \right)
      \cdot
      \ve{s}_k
      \label{eq:Breit_nucl_sm_1} \\
    & \quad
      -
      \frac{32 \pi}{3}
      \frac{\mu_{\tau_j} \mu_{\tau_k}}{c^2}
      \delta \left( \ve{r}_{jk} \right)
      \ve{s}_j \cdot \ve{s}_k
      \label{eq:Breit_nucl_sm_2} \\
    & \quad
      -
      \frac{e_{\tau_k} \hbar}{Mc^2}
      \left(
      2 \mu_{\tau_j}
      -
      \frac{e_{\tau_j} \hbar}{2 M}
      \right)
      \frac{1}{r_{jk}^3}
      \ve{s}_j
      \cdot
      \left( \ve{r}_{jk} \times \ve{p}_j \right)
      \label{eq:Breit_nucl_new_1} \\
    & \quad
      +
      \frac{e_{\tau_j} \hbar}{Mc^2}
      \left(
      2 \mu_{\tau_k}
      -
      \frac{e_{\tau_k} \hbar}{2 M}
      \right)
      \frac{1}{r_{jk}^3}
      \ve{s}_k
      \cdot
      \left( \ve{r}_{jk} \times \ve{p}_k \right)
      \label{eq:Breit_nucl_new_2} \\
    & \quad
      -
      \frac{\hbar \pi}{Mc^2}
      \left[
      e_{\tau_k} 
      \left(
      2 \mu_{\tau_j}
      -
      \frac{e_{\tau_j} \hbar}{2M}
      \right)
      +
      e_{\tau_j}
      \left(
      2 \mu_{\tau_k}
      -
      \frac{e_{\tau_k} \hbar}{2M}
      \right)
      \right]
      \delta \left( \ve{r}_{jk} \right)
      \label{eq:Breit_nucl_Darwin} \\
    & =
      -
      \delta_{p \tau_j}
      \delta_{p \tau_k}
      \frac{e^2}{2 M^2 c^2}
      \ve{p}_j 
      \cdot
      \left(
      \frac{1}{r_{jk}}
      +
      \frac{\ve{r}_{jk} \ve{r}_{jk}}{r_{jk}^3}
      \right)
      \cdot
      \ve{p}_k
      \notag \\
    & \quad
      -
      \frac{\hbar^2 e^2}{M^2 c^2}
      \frac{1}{r_{jk}^3}
      \ve{s}_j
      \cdot
      \left[
      \ve{r}_{jk}
      \times
      \left(
      \tilde{\mu}_{\tau_j} \delta_{p \tau_k}
      \ve{p}_k
      -
      \tilde{\mu}_{\tau_k} \delta_{p \tau_j}
      \ve{p}_j
      \right)
      \right]
      \notag \\
    & \quad
      -
      \tilde{\mu}_{\tau_j} \tilde{\mu}_{\tau_k}
      \frac{\hbar^2 e^2}{M^2 c^2}
      \ve{s}_j
      \cdot
      \left(
      \frac{3 \ve{r}_{jk} \ve{r}_{jk}}{r_{jk}^5}
      -
      \frac{1}{r_{jk}^3}
      \right)
      \cdot
      \ve{s}_k
      \notag \\
    & \quad
      -
      \tilde{\mu}_{\tau_j} \tilde{\mu}_{\tau_k}
      \frac{8 \pi \hbar^2 e^2}{3 M^2 c^2}
      \delta \left( \ve{r}_{jk} \right)
      \ve{s}_j \cdot \ve{s}_k
      \notag \\
    & \quad
      -
      \delta_{p \tau_k} 
      \frac{\hbar^2 e^2}{2 M^2 c^2}
      \left(
      2 \tilde{\mu}_{\tau_j}
      -
      \delta_{p \tau_j}
      \right)
      \frac{1}{r_{jk}^3}
      \ve{s}_j
      \cdot
      \left( \ve{r}_{jk} \times \ve{p}_j \right)
      \notag \\
    & \quad
      +
      \delta_{p \tau_j}
      \frac{\hbar^2 e^2}{2 M^2 c^2}
      \left(
      2 \tilde{\mu}_{\tau_k}
      -
      \delta_{p \tau_k}
      \right)
      \frac{1}{r_{jk}^3}
      \ve{s}_k
      \cdot
      \left( \ve{r}_{jk} \times \ve{p}_k \right)
      \notag \\
    & \quad
      -
      \frac{\pi \hbar^2 e^2}{2 M^2 c^2}
      \left[
      \delta_{p \tau_k} 
      \left(
      2 \tilde{\mu}_{\tau_j}
      -
      \delta_{p \tau_j}
      \right)
      +
      \delta_{p \tau_j}
      \left(
      2 \tilde{\mu}_{\tau_k}
      -
      \delta_{p \tau_k}
      \right)
      \right]
      \delta \left( \ve{r}_{jk} \right),
      \notag 
  \end{align}
\end{subequations}
where $ M \approx 938.919 \, \mathrm{MeV} / c^2 $ is the nucleon mass~\cite{
  Zyla2020Prog.Theor.Exp.Phys.2020_083C01},
$ \delta_{p \tau} $ is the Kronecker delta,
which is equal to $ 1 $ for $ \tau = p $ and $ 0 $ for $ \tau = n $,
$ e_{\tau} $ and $ \mu_{\tau} $ are the charge and magnetic moment of a nucleon $ \tau $,
and
$ \tilde{\mu}_{\tau} $ is the magnetic moment of the unit of the nuclear magneton, i.e.,~\cite{
  Zyla2020Prog.Theor.Exp.Phys.2020_083C01}
\begin{subequations}
  \begin{align}
    \mu_{\tau}
    & =
      \tilde{\mu}_{\tau}
      \frac{e \hbar}{2M}, \\
    \tilde{\mu}_p
    & =
      2.792 847 344 63, \\
    \tilde{\mu}_n
    & =
      -1.913 042 73.
  \end{align}
\end{subequations}
\par
If one assumes that these two particles are electrons instead of nucleons,
and
accordingly, $ m_e $, $ -e $, and $ \mu_e $, are used instead of $ M $, $ e_{\tau} $, and $ \mu_{\tau} $,
one can find that Eq.~\eqref{eq:Breit_nucl} is identical to Eq.~\eqref{eq:Breit_NR}.
Here, the relation between the $ g $-factor and the magnetic moment $ \mu_e $
\begin{equation}
  \mu_e
  =
  \frac{e \hbar}{4 m_e}
  g_e
\end{equation}
and $ g_e = 2 $ is used,
apart from the second term of Eq.~\eqref{eq:Breit_nucl_so},
in which the $ g $-factor for the orbital motion of electrons $ g_e = 1 $ is used.
In more detail,
the correspondence between Eqs.~\eqref{eq:Breit_NR} and \eqref{eq:Breit_nucl} is shown in Table~\ref{tab:Breit_electron}.
In the case of electronic systems,
Eqs.~\eqref{eq:Breit_nucl_new_1} and \eqref{eq:Breit_nucl_new_2} are cancelled out of each other.
\begin{table*}[tb]
  \centering
  \caption{Correspondence between Eqs.~\eqref{eq:Breit_NR} and \eqref{eq:Breit_nucl}.
    For convenience, correspondence to Eq.~(2) of Ref.~\cite{
      Itoh1965Rev.Mod.Phys.37_159}
    is also shown.}
  \label{tab:Breit_electron}
  \begin{tabular}{llll}
    \hline \hline
    \multicolumn{1}{l}{Eq.~\eqref{eq:Breit_nucl}} & \multicolumn{1}{l}{Eq.~\eqref{eq:Breit_NR}} & \multicolumn{1}{l}{Eq.~(2) of Ref.~\cite{Itoh1965Rev.Mod.Phys.37_159}} & \multicolumn{1}{l}{Origin} \\
    \hline
    Eq.~\eqref{eq:Breit_nucl_retar}  & Eq.~\eqref{eq:Breit_NR_retar}  & Eq.~(2b) & Retardation \\
    Eq.~\eqref{eq:Breit_nucl_so}     & Eq.~\eqref{eq:Breit_NR_so}     & Eqs.~(2c) and (2d) & Spin-orbit interaction \\
    Eq.~\eqref{eq:Breit_nucl_sm_1}   & Eq.~\eqref{eq:Breit_NR_sm_2}   & Eq.~(2e) & Spin-magnetic interaction (non-mutually penetrating) \\
    Eq.~\eqref{eq:Breit_nucl_sm_2}   & Eq.~\eqref{eq:Breit_NR_sm_1}   & Eq.~(2f) & Spin-magnetic interaction (mutually penetrating) \\
    Eq.~\eqref{eq:Breit_nucl_Darwin} & Eq.~\eqref{eq:Breit_NR_Darwin} & Eq.~(2g) & Darwin term \\
    \hline \hline
  \end{tabular}
\end{table*}
\subsection{Density functional theory for relativistic correction}
\label{subsec:DFT}
\par
In nuclear DFT, the ground-state energy is obtained by 
\begin{equation}
  \label{eq:DFT_gs}
  E_{\urm{gs}}
  =
  T_0
  +
  E_{\urm{nucl}} \left[ \rho_p^{\urm{gs}}, \rho_n^{\urm{gs}} \right]
  +
  E_{\urm{EM}} \left[ \rho_p^{\urm{gs}} \right], 
\end{equation}
where
$ T_0 $ is the Kohn-Sham kinetic energy, 
$ E_{\urm{nucl}} $ and $ E_{\urm{EM}} $ are EDFs of 
nuclear and EM interactions, respectively,
and
$ \rho^p_{\urm{gs}} $ and $ \rho^n_{\urm{gs}} $ are
the ground-state densities of protons and neutrons.
If the finite charge distributions of nucleons are considered,
$ E_{\urm{EM}} $ should depend on 
the ground-state charge distribution, instead of $ \rho_p^{\urm{gs}} $~\cite{
  Naito2020Phys.Rev.C101_064311}.
\par
If one only considers the Coulomb interaction to $ V_{\urm{EM}} $, 
$ E_{\urm{EM}} $ in LDA reads~\cite{
  Slater1951Phys.Rev.81_385}
\begin{subequations}
  \label{eq:CoulEDFall}
  \begin{align}
    E_{\urm{EM}} \left[ \rho \right]
    & = 
      E_{\urm{Coul}}^{\urm{H}} \left[ \rho \right]
      +
      E_{\urm{Coul}}^{\urm{x}} \left[ \rho \right], 
      \label{eq:CoulEDF} \\
    E_{\urm{Coul}}^{\urm{H}} \left[ \rho \right]
    & = 
      \frac{e^2}{2}
      \iint
      \frac{\rho \left( \ve{r} \right) \rho \left( \ve{r}' \right)}{\left| \ve{r} - \ve{r}' \right|}
      \, d \ve{r} \, d \ve{r}',
      \label{eq:CoulEDFH} \\
    E_{\urm{Coul}}^{\urm{x}} \left[ \rho \right]
    & =
      -
      \frac{3 e^2}{4}
      \left( \frac{3}{\pi} \right)^{1/3}
      \int
      \left[
      \rho \left( \ve{r} \right)
      \right]^{4/3}
      \, d \ve{r},
      \label{eq:CoulEDFx}
  \end{align}
\end{subequations}
where the density $ \rho $ is the electron density $ \rho_e $ for electronic systems
and
the proton density $ \rho_p $ for nuclear systems.
\par
Recently, an EDF for the Breit correction [Eq.~\eqref{eq:Breit_NR}] for electronic systems in the LDA
was developed in Refs.~\cite{
  Kenny1996Phys.Rev.Lett.77_1099,
  Naito2020J.Phys.B53_215002}.
In this paper, I consider the relativistic correction
[Eq.~\eqref{eq:Breit_nucl}]
in the nuclear DFT
using the idea of such an EDF and the correspondence shown in Table~\ref{tab:Breit_electron}.
This newly developed EDF enables one to perform the self-consistent calculation.
To isolate the relativistic correction,
I shall not consider the nucleon finite-size effects~\footnote{
  Effects of nucleon finite size were discussed in Refs.~\cite{
    Auerbach1983Phys.Rep.98_273,
    Naito2020Phys.Rev.C101_064311}.}.
\par
The Hartree and LDA exchange EDFs for the Breit correction in the electronic systems
[Eq.~\eqref{eq:Breit_NR}]
were proposed in Ref.~\cite{
  Naito2020J.Phys.B53_215002}
and Ref.~\cite{
  MacDonald1979J.Phys.C12_2977},
respectively.
Their forms are
\begin{subequations}
  \label{eq:RelEDF}
  \begin{align}
    E_{\urm{Breit}}^{\urm{H}} \left[ \rho \right]
    & =
      -
      \frac{\pi \hbar^2 e^2}{2 m_e^2 c^2}
      \int
      \left[
      \rho \left( \ve{r} \right)
      \right]^2
      \, d \ve{r}, 
      \label{eq:RelH} \\
    E_{\urm{Breit}}^{\urm{x}} \left[ \rho \right]
    & =
      \frac{3 \pi \hbar^2 e^2}{2 m_e^2 c^2}
      \int
      \left[
      \rho \left( \ve{r} \right)
      \right]^2
      \, d \ve{r},
      \label{eq:Relx}
  \end{align}
  respectively,
  and accordingly, the total correction is
  \begin{align}
    E_{\urm{Breit}}^{\urm{tot}} \left[ \rho \right]
    & =
      E_{\urm{Breit}}^{\urm{H}} \left[ \rho \right]
      +
      E_{\urm{Breit}}^{\urm{x}} \left[ \rho \right]
      \notag \\
    & = 
      \frac{\pi \hbar^2 e^2}{m_e^2 c^2}
      \int
      \left[
      \rho \left( \ve{r} \right)
      \right]^2
      \, d \ve{r}.
      \label{eq:Reltot} 
  \end{align}
\end{subequations}
The idea of the LDA is that the energy density is approximated to that of homogeneous systems;
hence,
contributions of Eqs.~\eqref{eq:Breit_NR_sm_1}, \eqref{eq:Breit_NR_sm_2}, and \eqref{eq:Breit_NR_so} vanish.
Consequently,
Eqs.~\eqref{eq:RelH} and \eqref{eq:Relx} contain only effects originating from Eqs.~\eqref{eq:Breit_NR_Darwin} and \eqref{eq:Breit_NR_retar}~\cite{
  Kenny1996Phys.Rev.Lett.77_1099}.
In addition, due to the symmetry, Eq.~\eqref{eq:RelH} does not contain the effect of the retardation [Eq.~\eqref{eq:Breit_NR_retar}].
\par
I shall construct an EDF for Eq.~\eqref{eq:Breit_nucl} in the LDA, using the knowledge of Eqs.~\eqref{eq:RelH} and \eqref{eq:Relx}.
Because of the idea of the LDA, the EDF contains only the effect of Eqs.~\eqref{eq:Breit_nucl_retar} and \eqref{eq:Breit_nucl_Darwin},
as in the electronic systems.
Note that the spin-orbit interaction originating from the magnetic form factors of nucleons
[Eqs.~\eqref{eq:Breit_nucl_so}, \eqref{eq:Breit_nucl_new_1}, and \eqref{eq:Breit_nucl_new_2}]
have been discussed in Refs.~\cite{
  Roca-Maza2018Phys.Rev.Lett.120_202501,
  Roca-Maza2018EPJWebConf.194_01002,
  Naito2020Phys.Rev.C101_064311};
the spin-magnetic interaction [Eqs.~\eqref{eq:Breit_nucl_sm_1} and \eqref{eq:Breit_nucl_sm_2}]
has not been considered yet for the self-consistent calculation
and
it cannot be considered in the LDA,
which is left for the future study.
\par
Then, I will construct the LDA EDF for Eq.~\eqref{eq:Breit_nucl}.
Referring the result on Eq.~\eqref{eq:RelEDF},
one can find that
Eq.~\eqref{eq:Breit_nucl_Darwin} contributes both the Hartree and the exchange terms,
whereas Eq.~\eqref{eq:Breit_nucl_retar} contributes the exchange term only.
\par
The Hartree and the exchange EDFs of Eq.~\eqref{eq:Breit_nucl_Darwin}, respectively, read
\begin{subequations}
  \label{eq:EDF_Darwin}
  \begin{align}
    E_{\urm{Darwin}}^{\urm{H}} \left[ \rho_p, \rho_n \right]
    & =
      \int
      \left\{
      \frac{C_{pp}}{2}
      \left[ \rho_p \left( \ve{r} \right) \right]^2
      +
      C_{pn}
      \rho_p \left( \ve{r} \right) 
      \rho_n \left( \ve{r} \right)
      +
      \frac{C_{nn}}{2}
      \left[ \rho_n \left( \ve{r} \right) \right]^2
      \right\}
      \, d \ve{r}
      \notag \\
    & = 
      \int
      \left\{
      \frac{C_{pp}}{2}
      \left[ \rho_p \left( \ve{r} \right) \right]^2
      +
      C_{pn}
      \rho_p \left( \ve{r} \right) 
      \rho_n \left( \ve{r} \right)
      \right\}
      \, d \ve{r}, 
      \label{eq:EDF_Darwin_H} \\
    E_{\urm{Darwin}}^{\urm{x}} \left[ \rho_p, \rho_n \right]
    & =
      -
      \int
      \left\{
      \frac{C_{pp}}{4}
      \left[ \rho_p \left( \ve{r} \right) \right]^2
      +
      \frac{C_{nn}}{4}
      \left[ \rho_p \left( \ve{r} \right) \right]^2
      \right\}
      \, d \ve{r} 
      \notag \\
    & = 
      -
      \int
      \frac{C_{pp}}{4}
      \left[ \rho_p \left( \ve{r} \right) \right]^2
      \, d \ve{r} ,
      \label{eq:EDF_Darwin_x}
  \end{align}
\end{subequations}
where
\begin{equation}
  C_{\tau_j \tau_k}
  =
  -
  \frac{\pi \hbar^2 e^2}{2 M^2 c^2}
  \left[
    \delta_{p \tau_k} 
    \left(
      2 \tilde{\mu}_{\tau_j}
      -
      \delta_{p \tau_j}
    \right)
    +
    \delta_{p \tau_j}
    \left(
      2 \tilde{\mu}_{\tau_k}
      -
      \delta_{p \tau_k}
    \right)
  \right]
\end{equation}
and accordingly
\begin{subequations}
  \begin{align}
    C_{pp}
    & = 
      -
      \frac{\pi \hbar^2 e^2}{M^2 c^2}
      \left(
      2 \tilde{\mu}_p
      -
      1
      \right), \\
    C_{pn}
    & = 
      -
      \frac{\pi \hbar^2 e^2}{M^2 c^2}
      \tilde{\mu}_n, \\
    C_{nn}
    & =
      0.
  \end{align}
\end{subequations}
\par
The exchange EDF for Eq.~\eqref{eq:Breit_nucl_retar} is constructed from Eq.~\eqref{eq:Relx}.
Equation~\eqref{eq:Relx} does not include only a contribution from Eq.~\eqref{eq:Breit_NR_retar} but also that from Eq.~\eqref{eq:Breit_NR_Darwin}.
Hence, the exchange EDF for Eq.~\eqref{eq:Breit_NR_retar} reads
\begin{equation}
  \frac{3 \pi \hbar^2 e^2}{2 m_e^2 c^2}
  \int
  \left[ \rho \left( \ve{r} \right) \right]^2
  \, d \ve{r}
  -
  \frac{1}{4}
  \frac{\pi \hbar^2 e^2}{m_e^2 c^2}
  \int
  \left[ \rho \left( \ve{r} \right) \right]^2
  \, d \ve{r}
  =
  \frac{5 \pi \hbar^2 e^2}{4 m_e^2 c^2}
  \int
  \left[ \rho \left( \ve{r} \right) \right]^2
  \, d \ve{r},
\end{equation}
since the Darwin EDF for electronic systems is
\begin{equation}
  -
  \frac{1}{4}
  \frac{\pi \hbar^2 e^2}{m_e c^2}
  \int
  \left[ \rho \left( \ve{r} \right) \right]^2
  \, d \ve{r}.
\end{equation}
To apply the nuclear systems, with replacing the electron mass $ m_e $ to the nucleon mass $ M $,
one obtains
\begin{equation}
  \label{eq:EDF_retar}
  E_{\urm{retar}}^{\urm{x}} \left[ \rho_p, \rho_n \right]
  =
  \frac{5 \pi \hbar^2 e^2}{4 M^2 c^2}
  \int
  \left[ \rho_p \left( \ve{r} \right) \right]^2
  \, d \ve{r}.
\end{equation}
\par
Combining Eqs.~\eqref{eq:EDF_Darwin} and \eqref{eq:EDF_retar},
the total EDF for the correction reads
\begin{align}
  & E_{\urm{Rel}}^{\urm{tot}} \left[ \rho_p, \rho_n \right]
    \notag \\
  & =
    \int
    \left\{
    \frac{C_{pp}}{2}
    \left[ \rho_p \left( \ve{r} \right) \right]^2
    +
    C_{pn}
    \rho_p \left( \ve{r} \right) 
    \rho_n \left( \ve{r} \right)
    \right\}
    \, d \ve{r}
    -
    \int
    \frac{C_{pp}}{4}
    \left[ \rho_p \left( \ve{r} \right) \right]^2
    \, d \ve{r}
    +
    \frac{5 \pi \hbar^2 e^2}{4 M^2 c^2}
    \int
    \left[ \rho_p \left( \ve{r} \right) \right]^2
    \, d \ve{r}
    \notag \\
  & =
    \left(
    \frac{C_{pp}}{2}
    -
    \frac{C_{pp}}{4}
    +
    \frac{5 \pi \hbar^2 e^2}{4 M^2 c^2}
    \right)
    \int
    \left[ \rho_p \left( \ve{r} \right) \right]^2
    \, d \ve{r}
    +
    C_{pn}
    \int
    \rho_p \left( \ve{r} \right) 
    \rho_n \left( \ve{r} \right)
    \, d \ve{r}
    \notag \\
  & =
    \left(
    3
    -
    \tilde{\mu}_p
    \right)
    \frac{\pi \hbar^2 e^2}{2 M^2 c^2}
    \int
    \left[ \rho_p \left( \ve{r} \right) \right]^2
    \, d \ve{r}
    -
    \tilde{\mu}_n
    \frac{\pi \hbar^2 e^2}{M^2 c^2}
    \int
    \rho_p \left( \ve{r} \right) 
    \rho_n \left( \ve{r} \right)
    \, d \ve{r}
    \label{eq:corr_EDF} \\
  & \approx
    \frac{\pi \hbar^2 e^2}{M^2 c^2}
    \int
    \left[
    0.1035
    \rho_p \left( \ve{r} \right)
    +
    1.913
    \rho_n \left( \ve{r} \right)
    \right]
    \rho_p \left( \ve{r} \right) 
    \, d \ve{r}.
    \label{eq:corr_EDF_est} 
\end{align}
\section{Analytical Estimation}
\label{sec:analytical}
\par
Before performing numerical many-body calculations with the relativistic correction,
I estimate contribution of the correction to the total energy with a simple model
and discuss the systematic behavior.
As in Ref.~\cite{
  Naito2020Phys.Rev.C101_064311},
I assume hard-sphere distributions for both proton and neutron distribution $ \rho_p $ and $ \rho_n $:
\begin{equation}
  \label{eq:hard_proton_density}
  \rho_{\tau} \left( r \right)
  =
  \begin{cases}
    \rho_0^{\tau}
    & r \le R_{\tau}, \\
    0
    & r > R_{\tau},
  \end{cases}
\end{equation}
where $ R_{\tau} $ is the proton ($ \tau = p $) or neutron ($ \tau = n $) radius defined by
\begin{equation}
  \label{eq:hard_proton_radius}
  R_{\tau}
  =
  \left(
    \frac{3 N_{\tau}}{4 \pi \rho_0^{\tau}}
  \right)^{1/3} 
\end{equation}
with $ N_p = Z $ and $ N_n = N $.
For simplicity,
the half of the saturation density of atomic nuclei
$ \rho_0^p = \rho_0^n \approx 0.08 \, \mathrm{fm}^{-3} $ is assumed.
With this density, one can estimate the Coulomb Hartree and exchange energies as~\cite{
  Naito2020Phys.Rev.C101_064311}
\begin{subequations}
  \label{eq:simple_C}
  \begin{align}
    E_{\urm{Coul}}^{\urm{H}}
    & =
      \frac{3e^2}{5}
      \left( \frac{4 \pi \rho_0^p}{3} \right)^{1/3}
      Z^{5/3}
      \notag \\
    & \simeq
      0.60 Z^{5/3} \, \mathrm{MeV},
      \label{eq:simple_CH} \\
    E_{\urm{Coul}}^{\urm{x}}
    & =
      -
      \frac{3e^2}{4}
      \left( \frac{3 \rho_0^p}{\pi} \right)^{1/3}
      Z
      \notag \\
    & \simeq
      - 0.46 Z \, \mathrm{MeV},
      \label{eq:simple_Cx}
  \end{align}
\end{subequations}
respectively.
\par
Since relations
\begin{subequations}
  \begin{align}
    \int
    \left[ \rho_p \left( \ve{r} \right) \right]^2
    \, d \ve{r}
    & =
      \frac{4 \pi R_p^3}{3}
      \left( \rho_0^p \right)^2
      \notag \\
    & =
      \rho_0^p Z, \\
    \int
    \rho_p \left( \ve{r} \right) 
    \rho_n \left( \ve{r} \right) 
    \, d \ve{r}
    & =
      \frac{4 \pi R_{<}^3}{3}
      \rho_p^0
      \rho_n^0
      \notag \\
    & =
      \begin{cases}
        \rho_n^0
        Z
        & \text{for $ R_p < R_n $}, \\
        \rho_p^0
        N
        & \text{for $ R_p > R_n $}, 
      \end{cases}
  \end{align}
\end{subequations}
hold with $ R_{<} = R_p $ for $ R_p < R_n $ and $ R_{<} = R_n $ for $ R_p > R_n $,
one, accordingly, obtains
\begin{subequations}
  \label{eq:estimate}
  \begin{align}
    E_{\urm{Darwin}}^{\urm{H}}
    & =
      \frac{C_{pp}}{2}
      \frac{4 \pi R_p^3}{3}
      \left( \rho_0^p \right)^2
      +
      C_{pn}
      \frac{4 \pi R_{<}^3}{3}
      \rho_p^0
      \rho_n^0
      \notag \\
    & = 
      -
      \frac{2 \pi^2 \hbar^2 e^2}{3 M^2 c^2}
      \left[
      \left(
      2 \tilde{\mu}_p
      -
      1
      \right)
      R_p^3
      \left( \rho_0^p \right)^2
      +
      2 
      \tilde{\mu}_n
      R_{<}^3
      \rho_p^0
      \rho_n^0
      \right], 
      \label{eq:estimate_DH} \\
    E_{\urm{Darwin}}^{\urm{x}}
    & =
      -
      \frac{C_{pp}}{4} 
      \frac{4 \pi R_p^3}{3}
      \left( \rho_0^p \right)^2
      \notag \\
    & =
      \frac{\pi^2 \hbar^2 e^2}{3 M^2 c^2}
      \left(
      2 \tilde{\mu}_p
      -
      1
      \right)
      R_p^3
      \left( \rho_0^p \right)^2
      \notag \\
    & = 
      \frac{\pi \hbar^2 e^2}{4 M^2 c^2}
      \left(
      2 \tilde{\mu}_p
      -
      1
      \right)
      \rho_0^p
      Z
      \notag \\
    & \simeq
      0.018 Z \, \mathrm{MeV}, 
      \label{eq:estimate_Dx} \\
    E_{\urm{retar}}^{\urm{x}}
    & = 
      \frac{5 \pi^2 \hbar^2 e^2}{3 M^2 c^2}
      R_p^3
      \left( \rho_0^p \right)^2
      \notag \\
    & = 
      \frac{5 \pi \hbar^2 e^2}{4 M^2 c^2}
      \rho_0^p
      Z
      \notag \\
    & \simeq 
      0.020 Z \, \mathrm{MeV},
      \label{eq:estimate_ret} \\
    E_{\urm{Rel}}^{\urm{tot}}
    & = 
      \frac{2 \pi^2 \hbar^2 e^2}{3 M^2 c^2}
      \left[
      \left(
      3
      -
      \tilde{\mu}_p
      \right)
      R_p^3
      \left( \rho_0^p \right)^2
      -
      2
      \tilde{\mu}_n
      R_{<}^3
      \rho_p^0
      \rho_n^0
      \right].
      \label{eq:estimate_tot} 
  \end{align}
\end{subequations}
In the case of $ R_p < R_n $,
Eqs.~\eqref{eq:estimate_DH} and \eqref{eq:estimate_tot} read 
\begin{subequations}
  \begin{align}
    E_{\urm{Darwin}}^{\urm{H}}
    & = 
      -
      \frac{2 \pi^2 \hbar^2 e^2}{3 M^2 c^2}
      \left[
      \left(
      2 \tilde{\mu}_p
      -
      1
      \right)
      \left( \rho_0^p \right)^2
      +
      2 
      \tilde{\mu}_n
      \rho_p^0
      \rho_n^0
      \right]
      R_p^3
      \notag \\
    & = 
      -
      \frac{\pi \hbar^2 e^2}{2 M^2 c^2}
      \left[
      \left(
      2 \tilde{\mu}_p
      -
      1
      \right)
      \rho_0^p
      +
      2 
      \tilde{\mu}_n
      \rho_n^0
      \right]
      Z
      \notag \\
    & \simeq
      - 0.006 Z \, \mathrm{MeV}, \\
    E_{\urm{Rel}}^{\urm{tot}}
    & = 
      \frac{2 \pi^2 \hbar^2 e^2}{3 M^2 c^2}
      \left[
      \left(
      3
      -
      \tilde{\mu}_p
      \right)
      R_p^3
      \left( \rho_0^p \right)^2
      -
      2
      \tilde{\mu}_n
      R_p^3
      \rho_p^0
      \rho_n^0
      \right]
      \notag \\
    & = 
      \frac{\pi \hbar^2 e^2}{2 M^2 c^2}
      \left[
      \left(
      3
      -
      \tilde{\mu}_p
      \right)
      \rho_0^p
      -
      2
      \tilde{\mu}_n
      \rho_n^0
      \right]
      Z
      \notag \\
    & \simeq 
      0.032 Z \, \mathrm{MeV} ; 
      \label{eq:estimate_tot_Z} 
  \end{align}
\end{subequations}
in the case of $ R_p > R_n $,
they read
\begin{subequations}
  \begin{align}
    E_{\urm{Darwin}}^{\urm{H}}
    & = 
      -
      \frac{2 \pi^2 \hbar^2 e^2}{3 M^2 c^2}
      \left[
      \left(
      2 \tilde{\mu}_p
      -
      1
      \right)
      R_p^3
      \left( \rho_0^p \right)^2
      +
      2 
      \tilde{\mu}_n
      R_n^3
      \rho_p^0
      \rho_n^0
      \right]
      \notag \\
    & = 
      -
      \frac{\pi \hbar^2 e^2}{2 M^2 c^2}
      \rho_0^p
      \left[
      \left(
      2 \tilde{\mu}_p
      -
      1
      \right)
      Z
      +
      2 
      \tilde{\mu}_n
      N
      \right]
      \notag \\
    & \simeq
      -0.037 Z
      +
      0.031 N 
      \, \mathrm{MeV}, \\
    E_{\urm{Rel}}^{\urm{tot}}
    & = 
      \frac{2 \pi^2 \hbar^2 e^2}{3 M^2 c^2}
      \left[
      \left(
      3
      -
      \tilde{\mu}_p
      \right)
      R_p^3
      \left( \rho_0^p \right)^2
      -
      2
      \tilde{\mu}_n
      R_n^3
      \rho_p^0
      \rho_n^0
      \right]
      \notag \\
    & = 
      \frac{\pi \hbar^2 e^2}{2 M^2 c^2}
      \rho_0^p 
      \left[
      \left(
      3
      -
      \tilde{\mu}_p
      \right)
      Z
      -
      2
      \tilde{\mu}_n
      N
      \right]
      \notag \\
    & \simeq
      0.002 Z
      +
      0.031 N \, \mathrm{MeV}.
      \label{eq:estimate_tot_N} 
  \end{align}
\end{subequations}
Comparing Eqs.~\eqref{eq:simple_Cx} and \eqref{eq:estimate_tot_Z},
one can find that the relativistic correction to the total energy $ \left| E_{\urm{Rel}}^{\urm{tot}} \right| $ is $ 15 $ times smaller than
the Coulomb exchange energy $ \left| E_{\urm{Coul}}^{\urm{x}} \right| $.
This means that although $ E_{\urm{Rel}}^{\urm{tot}} $ is small compared to the other contributions,
one can estimate that $ E_{\urm{Rel}}^{\urm{tot}} $ for $ \nuc{Pb}{208}{} $ may be around $ 2 \, \mathrm{MeV} $,
which is a non-negligible contribution compared to the desired accuracy of $ E_{\urm{nucl}} $
($ O \left( 100 \right) \, \mathrm{keV} $).
\par
In atomic physics, the vacuum polarization to the Coulomb potential formed by an atomic nucleus
is the next-leading order to the Breit correction,
i.e., the relativistic correction of electron-electron interaction~\cite{
  Eides2001Phys.Rep.342_63}.
On the other hand, in nuclear structure,
both the vacuum polarization and the relativistic correction
are corrections of the proton-proton Coulomb interaction.
Thus, it is worthwhile to compare them.
In Ref.~\cite{
  Naito2020Phys.Rev.C101_064311},
the vacuum polarization was taken into account in a nuclear structure calculation
using the Uehling potential~\cite{
  Uehling1935Phys.Rev.48_55,
  WayneFullerton1976Phys.Rev.A13_1283}.
The contribution of the vacuum polarization was estimated as~\cite{
  Naito2020Phys.Rev.C101_064311}
\begin{equation}
  \label{eq:simple_VP}
  E_{\urm{VP}}
  \simeq
  0.0047 Z^{5/3} \, \mathrm{MeV}.
\end{equation}
Hence, the relativistic correction and the vacuum polarization provide comparable contributions to the total energy in the light nuclei
(e.g., for $ Z = 20 $,
$ E_{\urm{Rel}}^{\urm{tot}} \approx 0.64 \, \mathrm{MeV} $
and
$ E_{\urm{VP}} \approx 0.69 \, \mathrm{MeV} $)
but
the latter dominates in heavy nuclei
(e.g., for $ Z = 100 $,
$ E_{\urm{Rel}}^{\urm{tot}} \approx 3.2 \, \mathrm{MeV} $
and
$ E_{\urm{VP}} \approx 10 \, \mathrm{MeV} $).
In the next section, I shall confirm this by performing numerical calculations.
\section{Skyrme Hartree--Fock Calculation}
\label{sec:SHF}
\par
To calculate the ground-state density distribution and energy numerically,
one needs to use a many-body calculation technique.
In this paper, the Skyrme Hartree--Fock method is used.
\par
The EDF for the correction [Eq.~\eqref{eq:corr_EDF}] is implemented to
the self-consistent Skyrme Hartree--Fock calculation code \textsc{skyrme\_rpa}~\cite{
  Colo2013Comput.Phys.Commun.184_142}.
In this work, doubly-magic nuclei are focused on;
thus, the pairing interaction is not considered,
and the spherical symmetry is assumed.
The calculations are performed with a box of $ 15 \, \mathrm{fm} $ with $ 0.1 \, \mathrm{fm} $ mesh.
The SAMi EDF~\cite{
  Roca-Maza2012Phys.Rev.C86_031306}
is adopted for the nuclear EDF.
\par
It is worthwhile to mention here that the non-relativistic Coulomb GGA exchange EDF,
more precisely, the modified Coulomb GGA-Perdew--Burke--Ernzerhof (PBE) EDF with $ \lambda = 1.25 $~\cite{
  Naito2019Phys.Rev.C99_024309},
reproduces the exact Coulomb Fock energy less than $ 100 \, \mathrm{keV} $ error,
which is much less than the relativistic correction, which I will discuss;
thus, in this paper, the exact Coulomb Fock energy is approximated to the Coulomb GGA exchange energy,
$ E_{\urm{Coul}}^{\urm{x-GGA}} \simeq E_{\urm{Coul}}^{\urm{x-exact}} $.
\par
First of all, I shall mention effects to nuclear radii.
Neither the proton nor the neutron radii are found to be changed more than $ 0.002 \, \mathrm{fm} $ due to the relativistic correction and the vacuum polarization.
Therefore, the effect of the correction on nuclear radii can be safely neglected.
\par
Next, the ground-state energies for doubly-magic nuclei
are shown in Table \ref{tab:tot_energy}.
The columns labelled ``Coulomb (NR-LDA)'' and ``Coulomb (R-LDA)'' show
results with
the non-relativistic Coulomb LDA [Eq.~\eqref{eq:CoulEDFall}]
and
the relativistic Coulomb LDA [Eqs.~\eqref{eq:CoulEDFall} and \eqref{eq:corr_EDF}], respectively.
For comparison, total energies calculated with the vacuum polarization on top of the relativistic Coulomb LDA 
and with the non-relativistic Coulomb GGA~\cite{
  Naito2019Phys.Rev.C99_024309}
are shown in a column labelled ``C.~(R-LDA) \& V.P.''
and ``Coulomb (NR-GGA)'', respectively.
Both the relativistic correction and the vacuum polarization make the atomic nuclei less bound.
The contribution of the former is around $ 0.42 \, \mathrm{MeV} $ in $ \nuc{Ca}{40}{} $ and $ 2.37 \, \mathrm{MeV} $ in $ \nuc{Pb}{208}{} $.
Although the relativistic correction changes $ \Delta E_{\urm{tot}} $ of $ \nuc{Ca}{48}{} $ and $ \nuc{Ni}{48}{} $ by $ 0.5 \, \mathrm{MeV} $,
the mass difference is not changed.
\par
It was shown in Refs.~\cite{
  Roca-Maza2016Phys.Rev.C94_044313,
  Naito2019Phys.Rev.C99_024309}
that the exact treatment or its simplified treatment called GGA
of the Coulomb exchange energy makes $ \nuc{Ca}{40}{} $ bound more 
by around $ 0.32 \, \mathrm{MeV} $
and
$ \nuc{Pb}{208}{} $ by around $ 0.74\, \mathrm{MeV} $,
which can be understood as the correction of the density gradient on the LDA.
Hence, the relativistic correction of the Coulomb interaction
is almost comparable with or even larger than the gradient correction 
but opposite direction to total energies,
as in the case of atoms~\cite{
  Naito2020J.Phys.B53_215002}.
Thus, if both contributions are considered simultaneously, 
even it is not obvious whether the total energy decreases or increases.
Indeed, this fact was pointed out in the context of atomic physics in Ref.~\cite{
  Naito2020J.Phys.B53_215002},
while no first-principles relativistic Coulomb GGA EDF is available,
which remains an open question.
\par
I also compare the relativistic correction with the vacuum polarization.
The vacuum polarization also makes the atomic nuclei less bound,
and contribution to the total energy
is approximately $ 0.4 \, \mathrm{MeV} $ in $ \nuc{Ca}{40}{} $ and $ 3.7 \, \mathrm{MeV} $ in $ \nuc{Pb}{208}{} $.
Thus, as discussed in Sec.~\ref{sec:analytical},
the vacuum polarization 
gives a comparable contribution to the total energy to the relativistic correction in light nuclei,
but it contributes more in heavy nuclei.
\par
\begin{figure}[tb]
  \centering
  \includegraphics[width=1.0\linewidth]{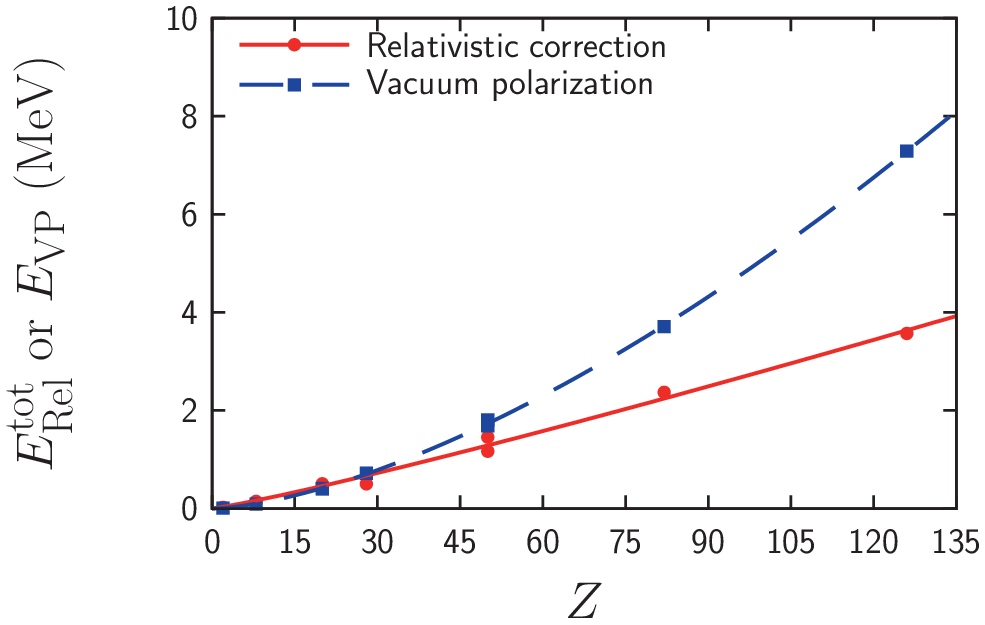}
  \caption{Contribution of relativistic correction and vacuum polarization to the total energies,
    $ E_{\urm{Rel}}^{\urm{tot}} $ and $ E_{\urm{VP}} $
    shown as functions of proton number $ Z $
    in circle and square points, respectively.
    Empirical fits with Eqs.~\eqref{eq:fit} are also shown
    by the solid and the dashed lines, respectively.}
  \label{fig:tot_energy}
\end{figure}

The systematic behavior of the relativistic correction and the vacuum polarization shall be discussed next. 
The contributions of the relativistic correction and the vacuum polarization can be seen as 
the difference of the second and the third columns and the third and the fourth columns in Table~\ref{tab:tot_energy}, respectively.
The proton number $ Z $ dependence of these contributions are shown in Fig.~\ref{fig:tot_energy}.
These energies are fitted with $ E = aZ^b $,
where $ a $ and $ b $ are adjustable parameters.
The results are
\begin{subequations}
  \label{eq:fit}
  \begin{align}
    E_{\urm{Rel}}^{\urm{tot}}
    & \simeq
      0.01660 Z^{1.12} \, \mathrm{MeV}, 
      \label{eq:fit_Breit} \\
    E_{\urm{VP}}
    & \simeq
      0.00398 Z^{1.55} \, \mathrm{MeV}, 
      \label{eq:fit_VP}
  \end{align}
\end{subequations}
which are consistent with Eqs.~\eqref{eq:estimate_tot_Z} and \eqref{eq:simple_VP}.
\begin{table*}[tb]
  \centering
  \caption{Ground-state energies calculated with the SAMi EDF~\cite{
      Roca-Maza2012Phys.Rev.C86_031306}
    shown in $ \mathrm{MeV} $.
    The columns labelled ``Coulomb (NR-LDA)'' and ``Coulomb (R-LDA)'' show
    results calculated with
    the non-relativistic Coulomb LDA [Eq.~\eqref{eq:CoulEDFall}]
    and
    the relativistic Coulomb LDA [Eqs.~\eqref{eq:CoulEDFall} and \eqref{eq:corr_EDF}], respectively.
    For comparison, total energies calculated with the vacuum polarization on top of the relativistic Coulomb LDA
    and with the non-relativistic Coulomb GGA
    are shown in columns labelled ``C.~(R-LDA) \& V.P.''
    and ``Coulomb (NR-GGA)'', respectively.}
  \label{tab:tot_energy}
  \begin{tabular}{rD{.}{.}{4}D{.}{.}{4}D{.}{.}{4}D{.}{.}{4}}
    \hline \hline
    \multicolumn{1}{c}{Nuclei} & \multicolumn{1}{c}{Coulomb (NR-LDA)} & \multicolumn{1}{c}{Coulomb (R-LDA)} & \multicolumn{1}{c}{C.~(R-LDA) \& V.P.} & \multicolumn{1}{c}{Coulomb (NR-GGA)} \\    
    \hline
    $ \nuc{He}{4}{} $    &   -27.5263 &   -27.5001 &   -27.4918 &   -27.6011 \\
    $ \nuc{O}{16}{} $    &  -130.4800 &  -130.3340 &  -130.2444 &  -130.6630 \\
    $ \nuc{Ca}{40}{} $   &  -347.0848 &  -346.6639 &  -346.2544 &  -347.4045 \\
    $ \nuc{Ca}{48}{} $   &  -415.6148 &  -415.1104 &  -414.7093 &  -415.9293 \\
    $ \nuc{Ni}{48}{} $   &  -352.6388 &  -352.1357 &  -351.4160 &  -353.0443 \\
    $ \nuc{Sn}{100}{} $  &  -811.6641 &  -810.4970 &  -808.6876 &  -812.2390 \\
    $ \nuc{Sn}{132}{} $  & -1103.0881 & -1101.6342 & -1099.9484 & -1103.6397 \\
    $ \nuc{Pb}{208}{} $  & -1636.6149 & -1634.2456 & -1630.5350 & -1637.3582 \\
    $ \nuc{126}{310}{} $ & -2131.4146 & -2127.8446 & -2120.5577 & -2132.3711 \\
    \hline \hline
  \end{tabular}
\end{table*}
\par
The breakdown of the relativistic correction [Eq.~\eqref{eq:corr_EDF}] is shown in Table~\ref{tab:rel_breakdown}.
The columns labelled $ E_{\urm{Darwin}}^{\urm{H}} $, $ E_{\urm{Darwin}}^{\urm{x}} $, and $ E_{\urm{retar}}^{\urm{x}} $ are
Darwin Hartree energy [Eq.~\eqref{eq:EDF_Darwin_H}], 
Darwin exchange one [Eq.~\eqref{eq:EDF_Darwin_x}], and
retardation exchange one [Eq.~\eqref{eq:EDF_retar}], respectively.
As shown, 
in $ N \le Z $ nuclei, $ E_{\urm{Darwin}}^{\urm{H}} < 0 $ holds,
while in $ N > Z $ nuclei, $ E_{\urm{Darwin}}^{\urm{H}} > 0 $ holds.
The Darwin Hartree energy reads
\begin{align}
  E_{\urm{Darwin}}^{\urm{H}}
  & =
    -
    \frac{\pi \hbar^2 e^2}{M^2 c^2}
    \int
    \left\{
    \frac{2 \tilde{\mu}_p - 1}{2}
    \rho_p \left( \ve{r} \right) 
    +
    \tilde{\mu}_n
    \rho_n \left( \ve{r} \right)
    \right\}
    \rho_p \left( \ve{r} \right) 
    \, d \ve{r}
    \notag \\
  & \simeq
    -
    \frac{\pi \hbar^2 e^2}{M^2 c^2}
    \int
    \left[
    2.292
    \rho_p \left( \ve{r} \right) 
    -
    1.913
    \rho_n \left( \ve{r} \right)
    \right]
    \rho_p \left( \ve{r} \right) 
    \, d \ve{r};
\end{align}
hence,
$ \left[
  2.292
  \rho_p \left( \ve{r} \right) 
  -
  1.913
  \rho_n \left( \ve{r} \right)
\right] $ 
determines the sign of $ E_{\urm{Darwin}}^{\urm{H}} $.
In $ N \le Z $ nuclei, $ \rho_p \left( \ve{r} \right) \gtrsim \rho_n \left( \ve{r} \right) $ holds
and, accordingly, it is positive (i.e., $ E_{\urm{Darwin}}^{\urm{H}} < 0 $),
whereas, in $ N > Z $ nuclei,
$ \rho_p \left( \ve{r} \right) \lesssim \rho_n \left( \ve{r} \right) $ holds
and accordingly, it can be negative (i.e., $ E_{\urm{Darwin}}^{\urm{H}} > 0 $)
in contrast to Eq.~\eqref{eq:estimate_DH}.
The Darwin and retardation exchange energies are also fitted with $ E = aZ^b $,
where the results are
\begin{subequations} 
  \label{eq:fit_rel}
  \begin{align}
    E_{\urm{Darwin}}^{\urm{x}}
    & \simeq
      0.01312 Z^{0.97} \, \mathrm{MeV},  
      \label{eq:fit_Dx} \\
    E_{\urm{retar}}^{\urm{x}}
    & \simeq
      0.01431 Z^{0.97} \, \mathrm{MeV},  
      \label{eq:fit_rx}
  \end{align}
\end{subequations}
which are consistent to Eqs.~\eqref{eq:estimate_Dx} and \eqref{eq:estimate_ret}.
\begin{table}[tb]
  \centering
  \caption{Breakdown of the relativistic correction [Eq.~\eqref{eq:corr_EDF}] shown in $ \mathrm{MeV} $.
    The columns labelled $ E_{\urm{Darwin}}^{\urm{H}} $, $ E_{\urm{Darwin}}^{\urm{x}} $, and $ E_{\urm{retar}}^{\urm{x}} $ are
    Darwin Hartree energy [Eq.~\eqref{eq:EDF_Darwin_H}], 
    Darwin exchange energy [Eq.~\eqref{eq:EDF_Darwin_x}], and
    retardation exchange energy [Eq.~\eqref{eq:EDF_retar}], respectively.} 
  \label{tab:rel_breakdown}
  \begin{tabular}{rD{.}{.}{4}D{.}{.}{4}D{.}{.}{4}}
    \hline \hline
    \multicolumn{1}{c}{Nuclei} & \multicolumn{1}{c}{$ E_{\urm{Darwin}}^{\urm{H}} $} & \multicolumn{1}{c}{$ E_{\urm{Darwin}}^{\urm{x}} $} & \multicolumn{1}{c}{$ E_{\urm{retar}}^{\urm{x}} $} \\
    \hline
    $ \nuc{He}{4}{} $    & -0.0048 & +0.0148 & +0.0162 \\
    $ \nuc{O}{16}{} $    & -0.0256 & +0.0821 & +0.0895 \\
    $ \nuc{Ca}{40}{} $   & -0.0704 & +0.2349 & +0.2562 \\
    $ \nuc{Ca}{48}{} $   & +0.0379 & +0.2230 & +0.2432 \\
    $ \nuc{Ni}{48}{} $   & -0.2429 & +0.3568 & +0.3890 \\
    $ \nuc{Sn}{100}{} $  & -0.1869 & +0.6475 & +0.7060 \\
    $ \nuc{Sn}{132}{} $  & +0.3026 & +0.5505 & +0.6002 \\
    $ \nuc{Pb}{208}{} $  & +0.4221 & +0.9311 & +1.0153 \\
    $ \nuc{126}{310}{} $ & +0.5492 & +1.4445 & +1.5750 \\
    \hline \hline
  \end{tabular}
\end{table}
\par
\begin{figure}[tb]
  \centering
  \includegraphics[width=1.0\linewidth]{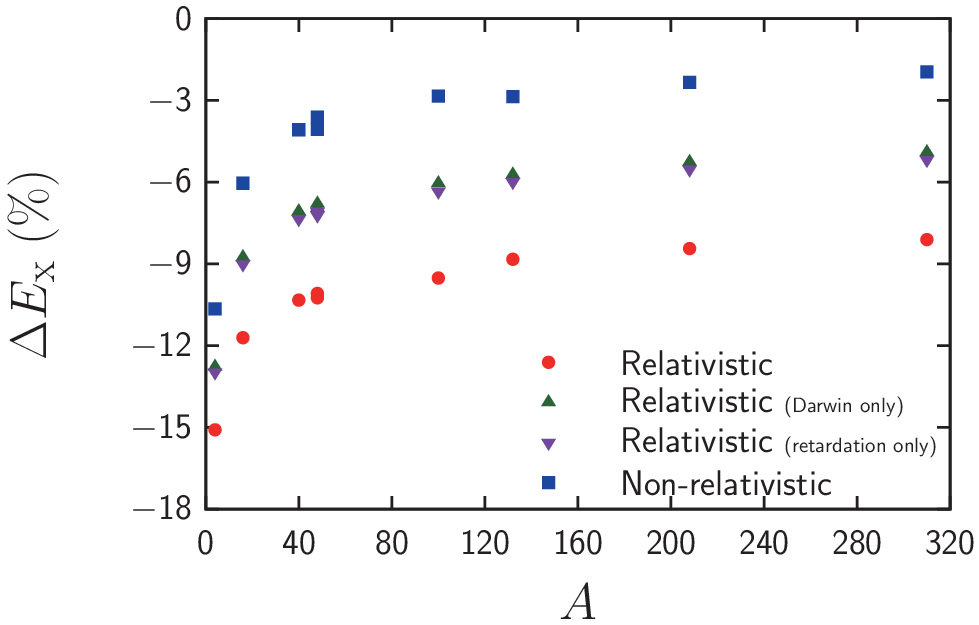}
  \caption{Ratio of the non-relativistic (square) or the relativistic (circle) Coulomb LDA exchange energies to the modified Coulomb GGA-PBE exchange energy with $ \lambda = 1.25 $,
    which reproduces the exact Coulomb Fock energy within enough accuracy,
    $ \Delta E_{\urm{x}} $ [Eq.~\eqref{eq:exchange}],
    shown as functions of mass number $ A $.
    The Darwin contribution only (up-triangle) and the retardation contribution only (down-triangle) are also shown.}
  \label{fig:exchange}
\end{figure}

The exchange contribution to the EM energy shall be further discussed.
The ratio of the Coulomb or relativistic exchange energy to the non-relativistic Coulomb GGA exchange energy
\begin{equation}
  \label{eq:exchange}
  \Delta E_{\urm{x}}
  =
  \frac{E^{\urm{x-LDA}} - E_{\urm{Coul}}^{\urm{x-GGA}}}{E_{\urm{Coul}}^{\urm{x-GGA}}}
  \simeq
  \frac{E^{\urm{x-LDA}} - E_{\urm{Coul}}^{\urm{x-exact}}}{E_{\urm{Coul}}^{\urm{x-exact}}}
\end{equation}
is shown in Fig.~\ref{fig:exchange}.
The non-relativistic and relativistic Coulomb LDA exchange energies, $ E^{\urm{x-LDA}} $,
are calculated by using $ E_{\urm{Coul}}^{\urm{x}} $ [Eq.~\eqref{eq:CoulEDFx}] 
and
$ E_{\urm{Coul}}^{\urm{x}} + E_{\urm{Darwin}}^{\urm{x}} + E_{\urm{retar}}^{\urm{x}} $ [Eqs.~\eqref{eq:EDF_Darwin_x} and \eqref{eq:EDF_retar}],
respectively.
It is seen that the absolute value of the relative change $ \left| \Delta E_{\urm{x}} \right| $
for the difference between the exact and LDA non-relativistic Coulomb exchange energy
gets smaller as $ A $ increases in light nuclei
and reaches almost constant
$ \Delta E_{\urm{x}} \approx -3 \, \% $ 
in heavy nuclei ($ A \gtrsim 100 $).
The absolute value of the relative change $ \left| \Delta E_{\urm{x}} \right| $
for the difference between the non-relativistic and relativistic LDA Coulomb exchange energy
behaves similarly,
while its value is larger 
and reaches almost constant
$ \Delta E_{\urm{x}} \approx -9 \, \% $ 
in heavy nuclei.
The Darwin contribution and the retardation one are almost the same size.
\par
This difference has been studied in the relativistic (covariant) DFT~\cite{
  Gu2013Phys.Rev.C87_041301,
  Niu2013Phys.Rev.C87_037301}.
The exact Coulomb exchange energy in this paper corresponds to Eq.~(4) in Ref.~\cite{
  Gu2013Phys.Rev.C87_041301},
which does not seem to include the relativistic correction.
They also show that the difference between the non-relativistic and relativistic Coulomb LDA exchange energies is approximately $ 1 \, \mathrm{MeV} $.
With considering such difference, $ E_{\urm{Coul}}^{\urm{x}} + E_{\urm{Rel}}^{\urm{x}} $ calculated in this work is almost consistent qualitatively with, 
but almost twice of Figs.~1 and 4 of Ref.~\cite{
  Gu2013Phys.Rev.C87_041301},
because Ref.~\cite{
  Gu2013Phys.Rev.C87_041301}
does not include $ E_{\urm{Darwin}}^{\urm{x}} $.
If one considers $ E_{\urm{Darwin}}^{\urm{x}} $ only,
this work is quantitatively consistent with Figs.~1 and 4 of Ref.~\cite{
  Gu2013Phys.Rev.C87_041301}.
\par
Lastly, a change of the single-particle energies is discussed,
for which $ \nuc{Pb}{208}{} $ is taken as an example.
The single-particle energies
calculated with
the non-relativistic and relativistic Coulomb LDA are shown in Table~\ref{tab:sp_Pb208}.
For comparison, those calculated with the vacuum polarization on top of the relativistic LDA are shown.
It is seen that the proton single-particle energies change by about $ 0.04 \, \mathrm{MeV} $ in inner shells
and $ 0.01 \, \mathrm{MeV} $ in outer shells,
due to the relativistic correction.
In contrast, the change due to the vacuum polarization does not depend on orbitals strongly,
and its value is about $ 0.1 \, \mathrm{MeV} $.
This difference is due to the nature of interaction;
the relativistic correction is the point-coupling interaction, while the vacuum polarization is finite range.
Since the Darwin Hartree term contributes to the neutron mean-field potential [see Eq.~\eqref{eq:corr_EDF_est}],
the neutron single-particle energies change by about $ 0.04 \, \mathrm{MeV} $ in inner shells
and $ 0.01 \, \mathrm{MeV} $ in outer shells,
due to the relativistic correction.
In contrast, the vacuum polarization changes neutron single-particle energies only by less than $ 0.01 \, \mathrm{MeV} $.
\begin{table}[tb]
  \centering
  \caption{Single-particle energies of $ \nuc{Pb}{208}{} $ shown in $ \mathrm{MeV} $ calculated with the SAMi EDF.}
  \label{tab:sp_Pb208}
  \begin{tabular}{lD{.}{.}{3}D{.}{.}{3}D{.}{.}{3}}
    \hline \hline
    \multicolumn{1}{c}{\multirow{2}{*}{States}} & \multicolumn{2}{c}{Coulomb} & \multicolumn{1}{c}{\multirow{2}{*}{C.~(R-LDA) \& V.P.}} \\
                                                & \multicolumn{1}{c}{NR-LDA} & \multicolumn{1}{c}{R-LDA} & \\
    \hline
    $ \pi 1s_{1/2} $ & -44.980 & -44.936 & -44.829 \\
    $ \pi 1p_{3/2} $ & -39.387 & -39.346 & -39.246 \\
    $ \pi 1p_{1/2} $ & -39.107 & -39.067 & -38.967 \\
    $ \pi 1d_{5/2} $ & -32.482 & -32.446 & -32.352 \\
    $ \pi 1d_{3/2} $ & -31.815 & -31.779 & -31.686 \\
    $ \pi 2s_{1/2} $ & -28.509 & -28.475 & -28.379 \\
    $ \pi 1f_{7/2} $ & -24.692 & -24.661 & -24.572 \\
    $ \pi 1f_{5/2} $ & -23.353 & -23.323 & -23.235 \\
    $ \pi 2p_{3/2} $ & -19.411 & -19.383 & -19.292 \\
    $ \pi 2p_{1/2} $ & -18.626 & -18.598 & -18.508 \\
    $ \pi 1g_{9/2} $ & -16.338 & -16.312 & -16.229 \\
    $ \pi 1g_{7/2} $ & -14.019 & -13.995 & -13.913 \\
    $ \pi 2d_{5/2} $ & -10.255 & -10.233 & -10.148 \\
    $ \pi 2d_{3/2} $ &  -8.846 &  -8.826 &  -8.742 \\
    $ \pi 3s_{1/2} $ &  -7.673 &  -7.653 &  -7.568 \\
    $ \pi 1h_{11/2}$ &  -7.663 &  -7.642 &  -7.563 \\
    \hline
    $ \nu 1s_{1/2} $ & -59.291 & -59.253 & -59.238 \\
    $ \nu 1p_{3/2} $ & -52.953 & -52.919 & -52.908 \\
    $ \nu 1p_{1/2} $ & -52.656 & -52.622 & -52.612 \\
    $ \nu 1d_{5/2} $ & -45.375 & -45.345 & -45.338 \\
    $ \nu 1d_{3/2} $ & -44.744 & -44.715 & -44.709 \\
    $ \nu 2s_{1/2} $ & -41.962 & -41.935 & -41.928 \\
    $ \nu 1f_{7/2} $ & -36.904 & -36.880 & -36.876 \\
    $ \nu 1f_{5/2} $ & -35.702 & -35.679 & -35.677 \\
    $ \nu 2p_{3/2} $ & -32.094 & -32.072 & -32.069 \\
    $ \nu 2p_{1/2} $ & -31.344 & -31.323 & -31.321 \\
    $ \nu 1g_{9/2} $ & -27.863 & -27.843 & -27.843 \\
    $ \nu 1g_{7/2} $ & -25.791 & -25.773 & -25.775 \\
    $ \nu 2d_{5/2} $ & -22.277 & -22.261 & -22.262 \\
    $ \nu 2d_{3/2} $ & -20.843 & -20.829 & -20.830 \\
    $ \nu 3s_{1/2} $ & -19.959 & -19.945 & -19.946 \\
    $ \nu 1h_{11/2}$ & -18.533 & -18.519 & -18.521 \\
    $ \nu 1h_{9/2} $ & -15.306 & -15.295 & -15.300 \\
    $ \nu 2f_{7/2} $ & -12.672 & -12.661 & -12.664 \\
    $ \nu 2f_{5/2} $ & -10.622 & -10.614 & -10.619 \\
    $ \nu 3p_{3/2} $ &  -9.859 &  -9.851 &  -9.854 \\
    $ \nu 3p_{1/2} $ &  -9.063 &  -9.055 &  -9.059 \\
    $ \nu 1i_{13/2}$ &  -9.150 &  -9.141 &  -9.145 \\
    \hline \hline
  \end{tabular}
\end{table}

%
%
%
%
%
\section{Conclusion and Perspectives}
\label{sec:conclusion}
\par
In this paper, the relativistic correction of the Coulomb interaction
was introduced in Skyrme Hartree--Fock calculations using the local density approximation (LDA).
The correction contains the Darwin term, the retardation, the spin-orbit interaction, and the spin-magnetic interaction,
while only the first and second terms were considered in this paper
because the remaining terms vanish in the LDA.
\par
It was found that the relativistic correction makes $ \nuc{Ca}{40}{} $ less bound by $ 0.4 \, \mathrm{MeV} $
and $ \nuc{Pb}{208}{} $ by  $ 2.4 \, \mathrm{MeV} $,
which are comparable with or even larger than  the difference between the exact Coulomb exchange (exact-Fock),
or its simplified treatment called the generalized gradient approximation (GGA),
and the LDA (Hartree--Fock--Slater approximation) Coulomb exchange energies~\cite{
  LeBloas2011Phys.Rev.C84_014310,
  Roca-Maza2016Phys.Rev.C94_044313,
  Naito2018Phys.Rev.C97_044319,
  Naito2019Phys.Rev.C99_024309},
but with the opposite sign.
Thus,
once both the exact treatment and the relativistic correction are considered at the same time,
even the sign of the total correction is not obvious,
as was pointed out also in the context of atomic physics~\cite{
  Naito2020J.Phys.B53_215002}.
In addition, since the desired accuracy of the nuclear EDF is $ 100 \, \mathrm{keV} $ order,
the relativistic correction is often non-negligible, as well as the correction due to the density gradient.
\par
The correction was also compared with the vacuum polarization.
Since the proton-number dependence of the vacuum polarization is stronger than that of the relativistic correction,
these two contributions to the total energy are comparable in light nuclei,
but the former dominates in heavy nuclei.
This behavior is in contrast to atoms,
in which the vacuum polarization is sub-dominant to the Breit correction~\cite{
  Eides2001Phys.Rep.342_63}.
Although the contribution of vacuum polarization to the total energy is usually tiny,
it is non-negligible to discuss mirror nuclei mass difference or the isobaric analog energy~\cite{
  Auerbach1983Phys.Rep.98_273,
  Roca-Maza2018Phys.Rev.Lett.120_202501,
  Naito2020Phys.Rev.C101_064311},
whereas the relativistic correction does not contribute to such properties.
\par
In this calculation, the spin-orbit and spin-magnetic interactions were not able to be considered
since they vanish in the LDA.
The former has been discussed in Refs.~\cite{
  Roca-Maza2018Phys.Rev.Lett.120_202501,
  Roca-Maza2018EPJWebConf.194_01002,
  Naito2020Phys.Rev.C101_064311},
while the latter has not been considered in the self-consistent calculation.
Consideration of the latter is left for a future task.
The relativistic Coulomb GGA functional is indispensable to achieve more accurate calculation in both atomic and nuclear structure calculations,
which is also left for a future perspective.
\par
The Coulomb correlation energy,
which originates from many-body correlations~\cite{
  Loewdin1955Phys.Rev.97_1474,
  Loewdin1955Phys.Rev.97_1490,
  Loewdin1955Phys.Rev.97_1509,
  Kato1957Commun.PureAppl.Math.10_151,
  Sinanoglu1968Phys.Rev.Lett.21_507},
is also an important and interesting topic.
In electronic systems, the Coulomb correlation energy can be defined without any ambiguity as the difference between the ``exact'' energy and the Hartree--Fock energy, which can be calculated by using the Monte Carlo calculation and some other many-body techniques~\cite{
  Ceperley1980Phys.Rev.Lett.45_566,
  Vosko1980Can.J.Phys.58_1200,
  Perdew1981Phys.Rev.B23_5048,
  Perdew1992Phys.Rev.B45_13244,
  Loos2012Int.J.QuantumChem.112_1712,
  Chachiyo2016J.Chem.Phys.145_021101,
  Karasiev2016J.Chem.Phys.145_157101,
  Yokota2021Phys.Rev.Research3_L012015}.
Reference~\cite{
  Kenny1996Phys.Rev.Lett.77_1099}
had derived the relativistic Coulomb exchange-correlation EDF in the LDA level,
while Ref.~\cite{
  Naito2020J.Phys.B53_215002}
discussed that the relativistic correction of the Coulomb correlation energy is almost negligible.
Meanwhile, in nuclear systems, there exists both the nuclear and Coulomb interaction;
hence, there is still ambiguity in defining the Coulomb correlation energy,
while there have been discussed in several works~\cite{
  Bulgac1996Nucl.Phys.A601_103,
  Bulgac1999Phys.Lett.B469_1,
  Bulgac1999Eur.Phys.J.A5_247,
  Naito2018Phys.Rev.C97_044319}.
Thus, it should be discussed whether the relativistic correction of the Coulomb correlation energy in nuclear systems is negligible,
while it is left for a future perspective.
\par
Once all the relevant corrections to the Coulomb energy are taken into account for the nuclear DFT calculation on top of the non-relativistic Coulomb LDA exchange EDF,
the correction may be more than $ 1 \, \mathrm{MeV} $ for heavy nuclei, such as $ \nuc{Pb}{208}{} $.
For instance, the nucleon finite-size effects and the vacuum polarization contribute to the binding energy of $ \nuc{Pb}{208}{} $, in total, in about $ 4 \, \mathrm{MeV} $~\cite{
  Naito2020Phys.Rev.C101_064311},
the relativistic correction discussed in this paper is about $ 2 \, \mathrm{MeV} $,
and
the Coulomb correlation energy may reach an order of $ 1 \, \mathrm{MeV} $~\cite{
  Bulgac1999Phys.Lett.B469_1}.
The current accuracy of nuclear EDFs is about $ 1 $--$ 5 \, \mathrm{MeV} $ in binding energies~\cite{
  Stoitsov2003Phys.Rev.C68_054312,
  Kortelainen2012Phys.Rev.C85_024304},
which is the same order as or even smaller than the Coulomb corrections.
Therefore, in order to achieve accurate calculation of binding energies using nuclear EDF quantitatively,
the nuclear EDF should be, in principle, refitted by considering all the relevant Coulomb corrections.

%
%
%
%
%
\section*{Acknowledgement}
The author thanks
Takumi Doi, Kenji Fukushima, 
Kouichi Hagino, Haozhao Liang, and Kenichi Yoshida for stimulating discussions.
The author also thanks Kouichi Hagino for variable comments on the manuscript.
The author acknowledges 
the JSPS Grant-in-Aid for JSPS Fellows under Grant No.~19J20543,
the JSPS Grant-in-Aid for Research Activity Start-up under Grant No.~22K20372,
the RIKEN iTHEMS program,
the RIKEN Special Postdoctoral Researchers Program,
and
the Science and Technology Hub Collaborative Research Program from RIKEN Cluster for Science, Technology and Innovation Hub (RCSTI).
The numerical calculations were performed on cluster computers at the RIKEN iTHEMS program.

%
%
%
\clearpage
%
%

%
\end{document}